\documentclass[11pt, a4paper]{article}
\usepackage{amsmath}
\usepackage{cite}
\usepackage{amsfonts}

\input{tcilatex}

\begin{document}

\title{Investigation of Volume Phase Transition from \\
the Different Properties of Particles}
\author{Yong Sun{\thanks{%
Email: ysun2002h@yahoo.com.cn}}}
\maketitle

\begin{abstract}
In this work, three different particle sizes: the static radius $R_{s}$,
hydrodynamic radius $R_{h}$ and apparent hydrodynamic radius $R_{h,app}$
obtained using the light scattering technique, are investigated for dilute
poly-disperse homogenous spherical particles with a simple assumption that
the hydrodynamic radius is in proportion to the static radius, when the
Rayleigh-Gans-Debye approximation is valid. The results show that the
expected values of the normalized time auto-correlation function of the
scattered light intensity $g^{\left( 2\right) }\left( \tau \right) $
calculated based on the static particle size information are consistent with
the experimental data. The volume phase transition is thus investigated
using the equilibrium swelling ratios of static radii and apparent
hydrodynamic radii respectively. The changes of the static particle size
information and apparent hydrodynamic radius as a function of temperature $T$
show the effects of the volume phase transition on optical properties and
the total influences of the volume phase transition on the optical,
hydrodynamic characteristics and size distribution of particles,
respectively. The effects of cross-linker on the volume phase transition are
also discussed.
\end{abstract}

\section{INTRODUCTION}

A great deal of information about particles in dispersion can be measured
using the light scattering technique. One of the main applications of the
light scattering technique is that measures the particle sizes. The static
light scattering technique (SLS) measures the size information from the
optical characteristics and the dynamic light scattering technique (DLS)
obtains the size information from both the optical and hydrodynamic features
of particles.

For a long time, the standard method of cumulants\cite{re1,re2,re3,re4} has
been used to measure the apparent hydrodynamic radius $R_{h,app}$ of
particles from the normalized time auto-correlation function of the
scattered light intensity $g^{\left( 2\right) }\left( \tau \right) $ with
the assistance of the Einstein-Stokes relation, where $\tau $ is the delay
time. The equilibrium swelling ratios\cite{re5,re6} of $R_{h,app}^{T}$ at
temperature $T$ over $R_{h,app}^{T_{0}}$ at temperature $T_{0}$ are used to
show the volume phase transition. Due to the fact that $R_{h,app}^{T}$ is
determined by the optical, hydrodynamic characteristics and size
distribution of particles and scattering vector\cite{re7}, the equilibrium
swelling ratios show the total changes of the optical, hydrodynamic
characteristics and size distribution of particles as a function of
temperature $T$. The treatment of SLS is simplified to the Zimm plot, Berry
plot or Guinier plot etc. to obtain the root mean-square radius of gyration $%
\left\langle R_{g}^{2}\right\rangle ^{1/2}$ and the molar mass of particles
provided that the particle sizes are small\cite{re4,re8}. In order to obtain
more information about the particles, people have explored the relationships
among the physical quantities measured using the SLS and DLS techniques
respectively. The measurements of the dimensionless shape parameter $\rho
=\left\langle R_{g}^{2}\right\rangle ^{1/2}/R_{h,app}$ \cite%
{re8,re9,re10,re11,re12} have been extensively used to infer the structures
of particles for a long time. In this judgement, it has an assumption that
the particle sizes measured from SLS and DLS are the same. However, the
particle sizes measured from SLS and DLS are different quantities\cite{re7}.
Exactly using the light scattering technique, three different sizes can be
measured for dilute poly-disperse homogenous spherical particles in
dispersion: one is a static radius $R_{s}$ measured from the optical
characteristics; the second is a hydrodynamic radius $R_{h}$ obtained from
the hydrodynamic features and the third is an apparent hydrodynamic radius $%
R_{h,app}$ determined by the optical, hydrodynamic characteristics and size
distribution of particles and scattering angle.

In this work, the three different particle sizes will be investigated using
Poly($N$-isopropylacrylamide)(PNIPAM) microgel samples with a simple
assumption that hydrodynamic radius $R_{h}$ is in proportion to the static
radius $R_{s}$. The results show that the expected values of $g^{\left(
2\right)} \left( \tau \right) $ calculated based on the particle size
information obtained using the static light technique are consistent with
the experimental data, the difference between the mean static radius and
apparent hydrodynamic radius is large and the difference between the mean
hydrodynamic radius and apparent hydrodynamic radius is influenced by the
particle size distribution. The volume phase transition is thus investigated
using the equilibrium swelling ratios of static radii and apparent
hydrodynamic radii respectively. The changes of the static particle size
information and apparent hydrodynamic radius as a function of temperature $T$
show the effects of the volume phase transition on optical properties and
the total influences of the volume phase transition on the optical,
hydrodynamic characteristics and size distribution of particles,
respectively. The effects of cross-linker on the volume phase transition are
also discussed.

\section{THEORY}

For dilute poly-disperse homogeneous spherical particles in dispersion where
the Rayleigh-Gans-Debye (RGD) approximation is valid, the normalized time
auto-correlation function of the electric field of the scattered light $%
g^{\left( 1\right) }\left( \tau \right) $ is given by

\begin{equation}
g^{\left( 1\right) }\left( \tau \right) =\frac{\int_{0}^{\infty
}R_{s}^{6}P\left( q,R_{s}\right) G\left( R_{s}\right) \exp \left(
-q^{2}D\tau \right) dR_{s}}{\int_{0}^{\infty }R_{s}^{6}P\left(
q,R_{s}\right) G\left( R_{s}\right) dR_{s}},  \label{Grhrs}
\end{equation}
where $R_{s}$ is the static radius, $D$ is the diffusion coefficient, $\ q=%
\frac{4\pi }{\lambda }n_{s}\sin \frac{\theta }{2}$ is the scattering vector, 
$\lambda $\ is the wavelength of the incident light in vacuo, $n_{s}$ is the
solvent refractive index, $\theta $ is the scattering angle, $G\left(
R_{s}\right) $ is the number distribution of particle sizes and the form
factor $P\left( q,R_{s}\right) $ is

\begin{equation}
P\left( q,R_{s}\right) =\frac{9}{q^{6}R_{s}^{6}}\left( \sin \left(
qR_{s}\right) -qR_{s}\cos \left( qR_{s}\right) \right) ^{2}.  \label{factor}
\end{equation}
In this work, the number distribution is chosen as a Gaussian distribution

\begin{equation}
G\left( R_{s};\left\langle R_{s}\right\rangle ,\sigma \right) =\frac{1}{
\sigma \sqrt{2\pi }}\exp \left( -\frac{1}{2}\left( \frac{R_{s}-\left\langle
R_{s}\right\rangle }{\sigma }\right) ^{2}\right) ,
\end{equation}
where $\left\langle R_{s}\right\rangle $ is the mean static radius and $%
\sigma $ is the standard deviation related to the mean static radius.

\noindent From the Einstein-Stokes relation

\begin{equation}
D=\frac{k_{B}T}{6\pi \eta _{0}R_{h}},
\end{equation}
where $\eta _{0}$, $k_{B}$ and $T$ are the viscosity of the solvent,
Boltzmann's constant and absolute temperature respectively, the hydrodynamic
radius $R_{h}$ can be obtained.

\noindent For simplicity, the relationship between the static and
hydrodynamic radii is assumed to be 
\begin{equation}
R_{h}=aR_{s},  \label{RsRh}
\end{equation}
where $a$ is a constant. From the Siegert relation between $g^{\left(
2\right) }\left( \tau \right) $ and $g^{\left( 1\right) }\left( \tau \right) 
$ \cite{re13}

\begin{equation}
g^{\left( 2\right) }\left( \tau \right) =1+\beta \left( g^{\left( 1\right)
}\right) ^{2},  \label{G1G2}
\end{equation}
the function between the static and dynamic light scattering is built and
the values of $g^{\left( 2\right) }\left( \tau \right) $ can be expected
based on the size information obtained using the SLS technique.

If the first cumulant is used to measure the apparent hydrodynamic radius $%
R_{h,app}$, the constant $a$ can be determined approximately using

\begin{equation}
a=\frac{R_{h,app}\int_{0}^{\infty }R_{s}^{6}P\left( q,R_{s}\right) G\left(
R_{s}\right) dR_{s}}{\int_{0}^{\infty }R_{s}^{7}P\left( q,R_{s}\right)
G\left( R_{s}\right) dR_{s}}.  \label{cona}
\end{equation}

\section{EXPERIMENT}

The SLS and DLS data were measured using the instrument built by ALV-Laser
Vertriebsgesellschaft m.b.H (Langen, Germany). It utilizes an ALV-5000
Multiple Tau Digital Correlator and a JDS Uniphase 1145P He-Ne laser to
provide a 23 mW vertically polarized laser at wavelength of 632.8 nm.

The PNIPAM microgel samples used in this work have been detailed before\cite%
{re7}. The four PNIPAM microgel samples PNIPAM-0, PNIPAM-1, PNIPAM-2 and
PNIPAM-5 were named according to the molar ratios $n_{B}/n_{N}$ of
cross-linker $N,N^{\prime}$-methylenebisacrylamide over $N$%
-isopropylacrylamide.

\section{RESULTS AND DISCUSSION}

The mean static radius $\left\langle R_{s}\right\rangle $ and standard
deviation $\sigma $ of the four PNIPAM microgel samples are measured from
the SLS data\cite{re7}. For PNIPAM-1, the mean static radii $\left\langle
R_{s}\right\rangle $, standard deviations $\sigma $ and $\chi ^{2}$\
obtained at different temperatures are listed in Table \ref{table1}.

\begin{table}[tbp]
\begin{center}
\begin{tabular}{|c|c|c|c|}
\hline
Temperature ($^{\text{o}}$C) & $\left\langle R_{s}\right\rangle $ (nm) & $%
\sigma $ (nm) & $\chi ^{2}$ \\ \hline
25 & 277.7$\pm $0.5 & 23.1$\pm $0.9 & 1.84 \\ \hline
27 & 267.1$\pm $0.5 & 23.1$\pm $0.8 & 2.50 \\ \hline
29 & 254.3$\pm $0.1 & 21.5$\pm $0.3 & 2.15 \\ \hline
31 & 224.8$\pm $0.9 & 30.6$\pm $0.9 & 3.31 \\ \hline
33 & 119.9$\pm 0.9$ & 19.8$\pm $0.6 & 3.16 \\ \hline
36 & 110.4$\pm $0.9 & 17.3$\pm $0.7 & 4.19 \\ \hline
40 & 111.7$\pm $0.9 & 14.8$\pm $0.8 & 2.73 \\ \hline
\end{tabular}
\makeatletter
\par
\makeatother 
\end{center}
\caption{The particle sizes obtained from SLS for PNIPAM-1 at different
temperatures.}
\label{table1}
\end{table}

The values of the apparent hydrodynamic radius at different scattering
angles were measured using the first cumulant analysis. In order to avoid
the consideration for the large values of $\chi ^{2}$, all the fit results
obtained using the first cumulant analysis are chosen under this condition $%
\chi ^{2}\leq 2$. For PNIPAM-1 at a temperature of 27 $^{\text{o}}$C, the
values of the apparent dynamic radius at different scattering angles are
listed in Table \ref{table2}. The ratios of the hydrodynamic radius over the
static radius calculated using Eq. \ref{cona} at different scattering angles
are also listed in Table \ref{table2}.

\begin{table}[tbp]
\begin{center}
\begin{tabular}{|c|c|c|}
\hline
Scattering Angle & $R_{h,app}$ (nm) & $a$ \\ \hline
30$^{\text{o}}$ & 329.$\pm $4. & 1.19 \\ \hline
35$^{\text{o}}$ & 331.1$\pm $0.7 & 1.21 \\ \hline
40$^{\text{o}}$ & 329.6$\pm $0.9 & 1.21 \\ \hline
45$^{\text{o}}$ & 329.8$\pm $0.5 & 1.22 \\ \hline
50$^{\text{o}}$ & 329.$\pm $1. & 1.22 \\ \hline
55$^{\text{o}}$ & 326.4$\pm $0.2 & 1.23 \\ \hline
60$^{\text{o}}$ & 327.$\pm $2. & 1.25 \\ \hline
65$^{\text{o}}$ & 323.$\pm $2. & 1.26 \\ \hline
\end{tabular}
\makeatletter
\par
\makeatother
\end{center}
\caption{Values of the apparent hydrodynamic radius and constant a for
PNIPAM-1 at different scattering angles and a temperature of 27 $^\mathrm o$%
C.}
\label{table2}
\end{table}

From the results shown in Table \ref{table2}, the values of $a$ at different
scattering angles are almost equal. If the value of constant $a$ was set to
1.21, the expected values of $g^{\left( 2\right) }\left( \tau \right) $
calculated based on the mean static radius $\left\langle R_{s}\right\rangle $
and standard deviation $\sigma $ using Eqs. \ref{Grhrs}, \ref{RsRh} and \ref%
{G1G2} were compared with the experimental data measured at scattering
angles 30$^{\text{o}}$, 45$^{\text{o}}$ and 65$^{\text{o}}$, respectively.
The results are shown in Fig. 1. The expected values are consistent with the
experimental data. With the constant $a$: 1.21, the mean hydrodynamic radius
323.2$\pm $0.6 nm can be obtained and this value only has little different
from the values of the apparent hydrodynamic radius measured using the first
cumulant at different scattering angles.

From the results shown in Table \ref{table1}, at a temperature of 33 $^{%
\text{o}}$C, the particle size distribution of PNIPAM-1 is wide. Its
corresponding values of the apparent hydrodynamic radius are listed in Table %
\ref{table3}. The ratios of the hydrodynamic radius over the static radius
calculated using Eq. \ref{cona} are also listed in Table \ref{table3}. The
values of $a$ at different scattering angles are almost equal. If the value
of constant $a$ was set to 1.55, the expected values of $g^{\left( 2\right)
}\left( \tau \right) $ calculated based on $\left\langle R_{s}\right\rangle $
and $\sigma $ using Eqs. \ref{Grhrs}, \ref{RsRh} and \ref{G1G2} were
compared with the experimental data measured at scattering angles 30$^{\text{%
o}}$, 60$^{\text{o}}$ and 90$^{\text{o}}$, respectively. The results are
shown in Fig. 2. The expected values are consistent with the experimental
data. With the constant $a$: 1.55, the mean hydrodynamic radius 186.$\pm $1.
nm is obtained and the difference between the mean hydrodynamic radius and
apparent hydrodynamic radius is large.

\begin{table}[tbp]
\begin{center}
\begin{tabular}{|c|c|c|}
\hline
Scattering Angle & $R_{h,app}$ (nm) & $a$ \\ \hline
30$^{\text{o}}$ & 212.1$\pm $0.6 & 1.55 \\ \hline
35$^{\text{o}}$ & 208.3$\pm $0.2 & 1.53 \\ \hline
40$^{\text{o}}$ & 208.7$\pm $0.6 & 1.54 \\ \hline
45$^{\text{o}}$ & 207.3$\pm $0.3 & 1.53 \\ \hline
50$^{\text{o}}$ & 206.7$\pm $0.4 & 1.53 \\ \hline
55$^{\text{o}}$ & 206.$\pm $1. & 1.53 \\ \hline
60$^{\text{o}}$ & 205.$\pm $1. & 1.53 \\ \hline
65$^{\text{o}}$ & 204.9$\pm $0.6 & 1.54 \\ \hline
70$^{\text{o}}$ & 205.$\pm $1. & 1.55 \\ \hline
75$^{\text{o}}$ & 205.$\pm $1. & 1.56 \\ \hline
80$^{\text{o}}$ & 205.$\pm $1. & 1.56 \\ \hline
85$^{\text{o}}$ & 205.$\pm $1. & 1.57 \\ \hline
90$^{\text{o}}$ & 203.2$\pm $0.7 & 1.57 \\ \hline
95$^{\text{o}}$ & 204.$\pm $1. & 1.58 \\ \hline
\end{tabular}
\makeatletter
\par
\makeatother
\end{center}
\caption{Values of the apparent hydrodynamic radius and constant a for
PNIPAM-1 at different scattering angles and a temperature of 33 $^\mathrm o$%
C.}
\label{table3}
\end{table}

From the results above, three different particle sizes can be measured using
the light scattering technique. In general, the values of the constant $a$
and apparent hydrodynamic radius $R_{h,app}$ vary with the scattering angle.
In order to compare the three different particle sizes conveniently, all the
mean hydrodynamic and apparent hydrodynamic radii were obtained at a
scattering angle of 30$^{\text{o}}$. All the three particle sizes measured
at different temperatures for PNIPAM-1 are shown in Fig. 3. The figure shows
that the difference between the mean static radius and apparent hydrodynamic
radius is large and the difference between the mean hydrodynamic radius and
apparent hydrodynamic radius is influenced by the particle size
distribution. The figure also shows that when the temperature nears the
phase transition temperature, all the radii have a sharp change and the
volumes of PNIPAM-1 collapse.

The three different particle sizes represent the different characteristics
of particles. When temperature changes from 25 $^{\text{o}}$C to 40 $^{\text{%
o}}$C, the property of the PNIPAM microgel samples changes from being
hydrophilic to hydrophobic. The volumes of PNIPAM microgel particles
collapse. It is possible that this change makes the different influences on
the optical and hydrodynamic characteristics of particles. In order to show
the effects of the volume phase transition, the ratios $R_{h,app}^{T}/\left%
\langle R_{s}^{T}\right\rangle $ and $\left\langle R_{h}^{T}\right\rangle
/\left\langle R_{s}^{T}\right\rangle $ as a function of temperature $T$ are
shown in Figs. 4a and 4b, respectively.

Figure 4 shows that the effects of the phase volume transition on the
optical and hydrodynamic characteristics of particles are different. The
differences are also influenced by the $N,N^{\prime }$%
-methylenebisacrylamide content. When the temperature nears the phase
transition temperature, the values of the ratios $R_{h,app}^{T}/\left\langle
R_{s}^{T}\right\rangle $ and $\left\langle R_{h}^{T}\right\rangle
/\left\langle R_{s}^{T}\right\rangle $ become larger and peaks emerge.
Figure 4a shows that the lower the crosslinker content, the higher the peak.
Figure 4b shows that the effects of the crosslinker content is not so
striking. From the static particle size distributions of the four PNIPAM
microgel samples, the particle size distributions are narrow both below and
above the phase transition and are wide near the phase transition. The
particle size distribution is also becoming narrow when the $N,N^{\prime }$%
-methylenebisacrylamide content is increased in the vicinity of the phase
transition temperature. From the theoretical analysis of cumulants, the
apparent hydrodynamic radius is obtained from averaging the term $\exp
\left( -q^{2}D\tau \right) $ in static particle size distribution $G\left(
R_{s}\right) $ with the weight $R_{s}^{6}P\left( q,R_{s}\right) $, where $%
\exp \left( -q^{2}D\tau \right) $ represents the hydrodynamic features of
particles. For the mono-disperse particle systems, since the effects of
scattered intensity are cancelled, the apparent hydrodynamic radius is equal
to the hydrodynamic radius. For poly-disperse particle systems, the apparent
hydrodynamic radius is different from the mean hydrodynamic radius and is
also determined by the particle size distribution\cite{re7}. Comparing Figs.
4a with 4b, the difference between the apparent hydrodynamic radius and mean
hydrodynamic radius is influenced obviously by the distribution width.

Since the PNIPAM microgels possess the temperature sensitivity in the
temperature range 15 $^{\text{o}}$C - 50 $^{\text{o}}$C, a few authors\cite%
{re12,re13} used the equilibrium swelling ratios $%
R_{h,app}^{T}/R_{h,app}^{T_{0}}$ of the apparent hydrodynamic radius at
temperature $T$ over that at temperature $T_{0}$ to show the volume phase
transition. For the four PNIPAM microgel samples, the volume phase
transition in the temperature range 25 $^{\text{o}}$C - 40 $^{\text{o}}$C is
shown in Fig. 5 using the equilibrium swelling ratios of the mean static
radii and the apparent hydrodynamic radii, respectively. All radii are
compared to that measured at a temperature of 40 $^{\text{o}}$C. The ratios $%
\left\langle R_{s}^{T}\right\rangle /\left\langle
R_{s}^{40^{o}C}\right\rangle $ and $R_{h,app}^{T}/R_{h,app}^{40^{o}C}$ are
shown in Fig. 5a and 5b respectively. Figure 5a shows the volume phase
transition is investigated using the optical properties and Figure. 5b shows
the volume phase transition is investigated using the optical, hydrodynamic
characteristics and size distribution of particles together.

From the material characteristics, the materials of PNIPAM possess the
temperature sensitivity. If adding the $N,N^{\prime }$%
-methylenebisacrylamide, the temperature sensitivity of PNIPAM microgels is
influenced by the content of the $N,N^{\prime }$-methylenebisacrylamide
which does not possess the temperature sensitivity. If the content of the $%
N,N^{\prime }$ -methylenebisacrylamide continues to increase, the
temperature sensitivity of PNIPAM microgels is becoming weak. Figure 5 shows
clearly the feature. The phase transition of PNIPAM microgels, indicated as
the ratios $\left\langle R_{s}^{T}\right\rangle /\left\langle
R_{s}^{40^{o}C}\right\rangle $ or $R_{h,app}^{T}/R_{h,app}^{40^{o}C}$ as a
function of $T$, becomes less sharp and occurs in a broader $T$ range as the 
$N,N^{\prime }$-methylenebisacrylamide content is increased.

In general, from the SLS data the root mean-square radius of gyration $%
\left\langle R_{g}^{2}\right\rangle ^{1/2}$ is obtained provided that the
particle sizes are small and from DLS data the hydrodynamic radius is got.
The dimensionless shape parameter $\rho $, the ratio of $\left\langle
R_{g}^{2}\right\rangle ^{1/2}$ to the hydrodynamic radius, has been
extensively used to infer the structure of particles\cite{re8,re9,re10} and
the change of $\rho $ as a function of temperature $T$ during the volume
phase transition range has been explored by a few authors\cite{re11,re12}.
Because the static and hydrodynamic radii are different quantities and
particles have a size distribution, $\left\langle R_{g}^{2}\right\rangle
^{1/2}/\left\langle R_{s}^{T}\right\rangle $ is determined not only by the
structure of particles but also the particle size distribution and $\rho $
includes not only the relationship between $\left\langle
R_{g}^{2}\right\rangle ^{1/2}$ and $\left\langle R_{s}^{T}\right\rangle $
but also the function between $\left\langle R_{s}^{T}\right\rangle $ and $%
R_{h,app}^{T}$. Since the effects of the volume phase transition on the
optical and hydrodynamic characteristics of particles are different, the
dimensionless parameters $\left\langle R_{g}^{2}\right\rangle
^{1/2}/\left\langle R_{s}^{T}\right\rangle $ and $\rho $ as a function of
temperature $T$ are shown in Figs. 6a and 6b, respectively. $\left\langle
R_{g}^{2}\right\rangle ^{1/2}$ was obtained using the relationship between
the $\left\langle R_{g}^{2}\right\rangle ^{1/2}$ and static size information%
\cite{re7}. As shown in Fig. 6a, the dimensionless parameter $\left\langle
R_{g}^{2}\right\rangle ^{1/2}/\left\langle R_{s}^{T}\right\rangle $ is
larger than 0.775 due to the effects of the particles size distribution.
When the temperature nears the phase transition temperature, the values of
the ratios $\left\langle R_{g}^{2}\right\rangle ^{1/2}/\left\langle
R_{s}^{T}\right\rangle $ become larger and peaks emerge. Since the lager the
value, the wider the particle size distribution, peaks show that the
particle size distribution become wide when the temperature nears the volume
phase transition temperature. $\rho $ has a very interesting change as a
function of temperature $T$ as shown in Fig. 6b. When the temperature nears
the phase transition temperature, the values of the ratios $\rho $ become
smaller and minimums emerge. Since the accurate relationship between the
optical and hydrodynamic features of particles has not been understood, the
meanings of $\rho $ need to be discussed further.

\section{CONCLUSION}

Using the light scattering technique, three different particle sizes can be
measured. The static radius represents the optical characteristics, the
hydrodynamic radius shows the hydrodynamic features and the apparent
hydrodynamic radius is determined by the optical, hydrodynamic
characteristics and size distribution of particles and scattering angle. The
difference between the mean static radius and apparent hydrodynamic radius
is large and the difference between the mean hydrodynamic radius and
apparent hydrodynamic radius is influenced by the particle size
distribution. The changes of the static particle size information and
apparent hydrodynamic radius as a function of temperature $T$ show the
effects of the volume phase transition on optical properties and the total
influences of the volume phase transition on the optical, hydrodynamic
characteristics and size distribution of particles, respectively. The
effects of the phase volume transition on the optical and hydrodynamic
characteristics of particles are different. The differences are also
influenced by the $N,N^{\prime }$-methylenebisacrylamide content. With the
static radius, the effects of the phase volume transition on the optical
properties have been explored. In order to explore the effects of the phase
volume transition on the hydrodynamic characteristics of particles and what $%
\rho $ means, the accurate relationship between the optical and hydrodynamic
features of particles must be understood.

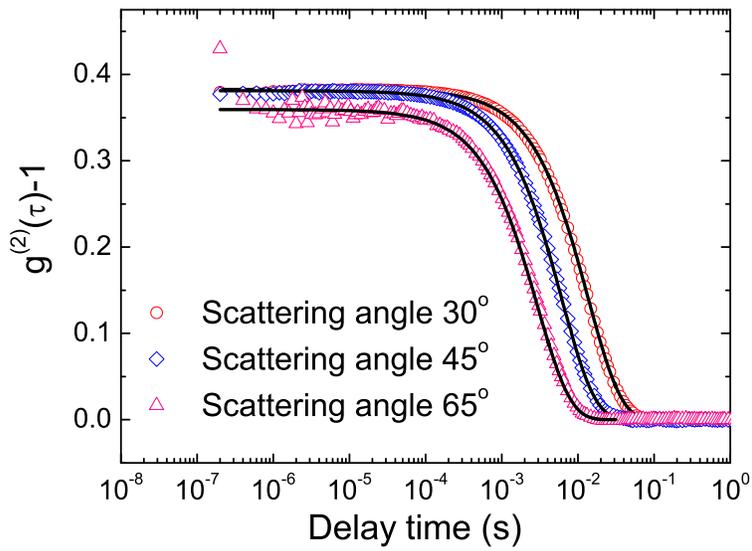
Fig. 1. The experimental data and expected values of $g^{\left( 2\right)
}\left( \tau \right) $ for PNIPAM-1 at a temperature of 27 $^{\text{o}}$C.
The symbols show the experimental data and the lines show the expected
values calculated under the simple assumption $R_{h}=1.21R_{s}$.

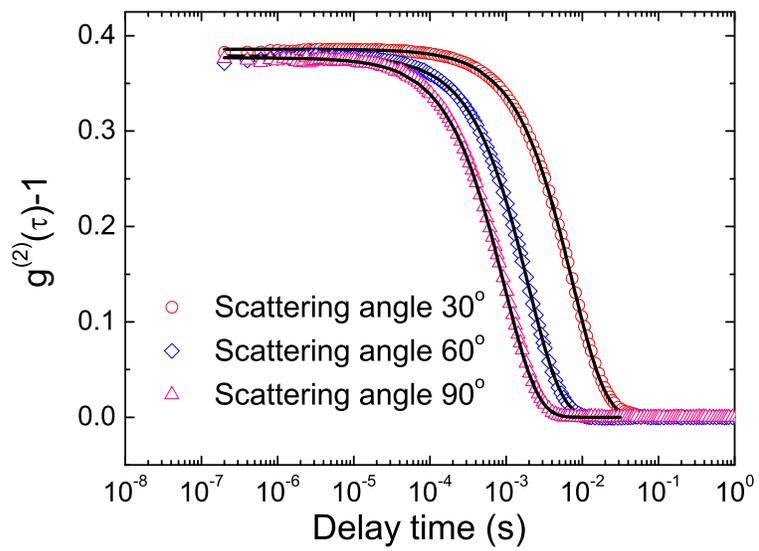
Fig. 2. The experimental data and expected values of $g^{\left( 2\right)
}\left( \tau \right) $ for PNIPAM-1 at a temperature of 33 $^{\text{o}}$C.
The symbols show the experimental data and the lines show the expected
values calculated under the simple assumption $R_{h}=1.55R_{s}$.

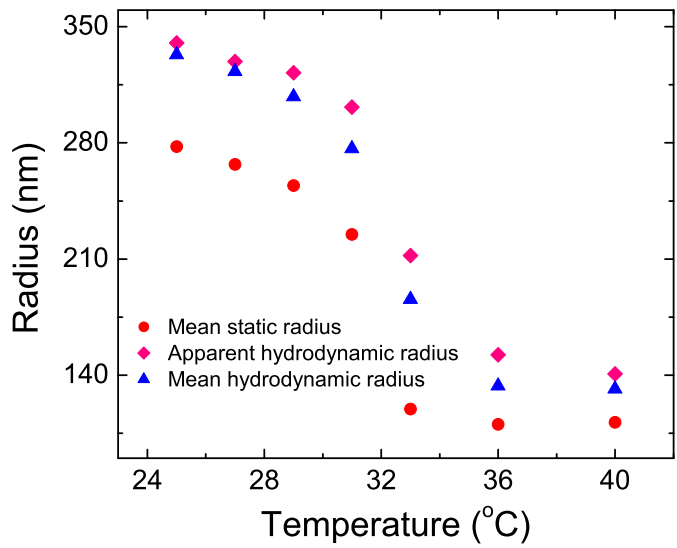
Fig. 3. Values of the apparent hydrodynamic radii, mean hydrodynamic radii
obtained at a scattering angle of 30$^{\text{o}}$ and mean static radii for
PNIPAM-1 at different temperatures.

Fig. 4. Ratios obtained from the three different particle sizes measured
using light scattering technique. a). The ratios $R_{h,app}^{T}/\left\langle
R_{s}^{T}\right\rangle $ of the apparent hydrodynamic radius over the mean
static radius are shown. b). The ratios $\left\langle R_{h}^{T}\right\rangle
/\left\langle R_{s}^{T}\right\rangle $ of the mean hydrodynamic radius over
the mean static radius are shown for PNIPAM-0, PNIPAM-1, PNIPAM-2 and
PNIPAM-5 in the volume phase transition temperature range from 25 $^{\text{o}%
}$C to 40 $^{\text{o}}$C.

Fig. 5. The volume phase transition of PNIPAM-0, PNIPAM-1, PNIPAM-2 and
PNIPAM-5. The phase transition is shown using the ratios of the mean static
radius $\left\langle R_{s}^{T}\right\rangle $ at temperature $T$ to that $%
\left\langle R_{s}^{40^{o}C}\right\rangle $ at 40 $^{\text{o}}$C (a) and the
ratios of the apparent hydrodynamic radius $R_{h,app}^{T}$ at temperature $T$
to that $R_{h,app}^{40^{o}C}$ at 40 $^{\text{o}}$C (b).

Fig. 6. The effects of the volume phase transition on the dimensionless
parameters $\left\langle R_{g}^{2}\right\rangle ^{1/2}/\left\langle
R_{s}^{T}\right\rangle $ and $\rho $. The results of $\left\langle
R_{g}^{2}\right\rangle ^{1/2}/\left\langle R_{s}^{T}\right\rangle $ and $%
\rho $ as a function of temperature $T$ are shown in a and b, respectively.

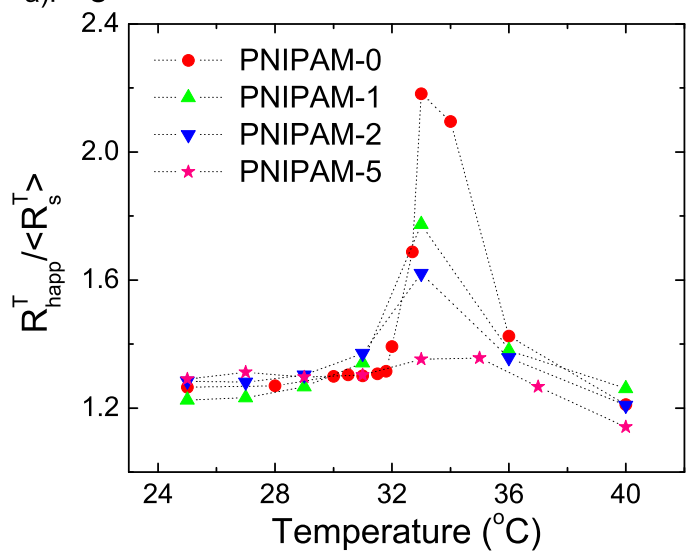
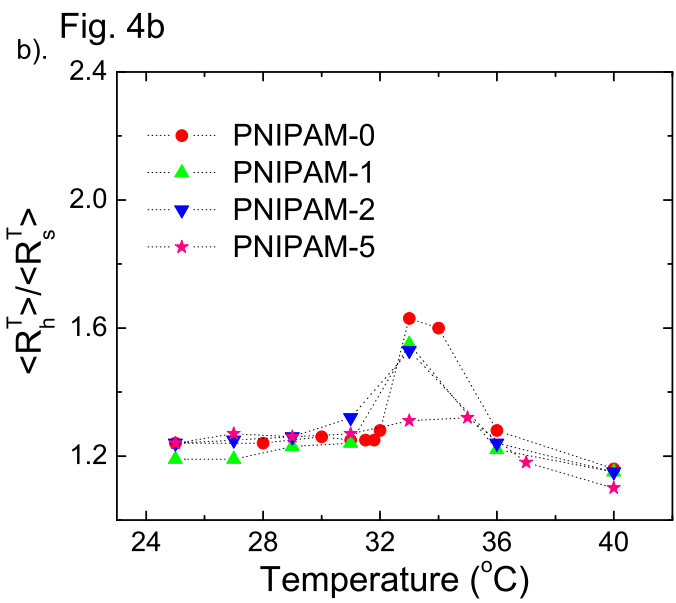
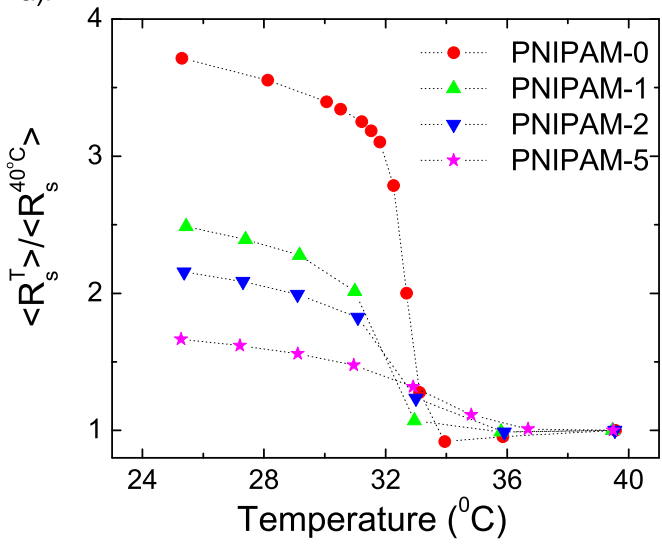
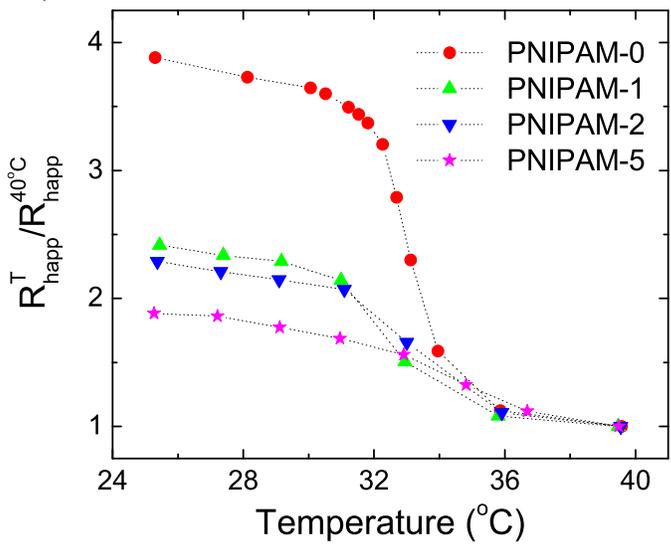
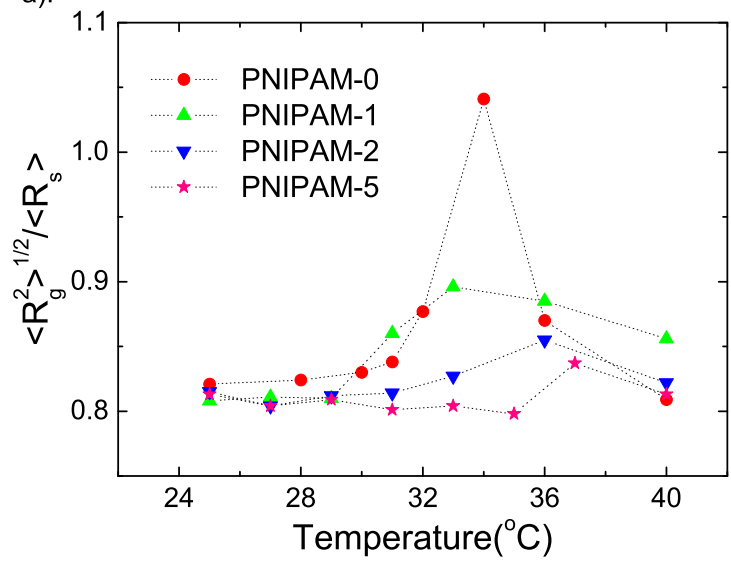
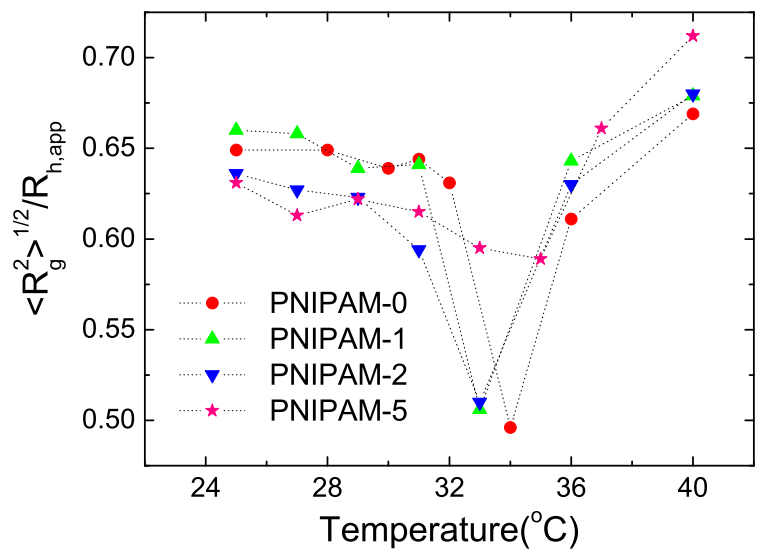
\end{document}